\documentclass[sigconf]{acmart}
\usepackage{graphicx,xcolor}
\usepackage[T1]{fontenc}
\usepackage{xurl}
\usepackage{cleveref}
\usepackage{tabularx}
\usepackage{listings}

\usepackage[labelformat=simple]{subcaption} 

\crefname{figure}{Figure}{Figure}
\crefname{table}{Table}{Table}
\crefname{section}{Section}{Sections}
\crefname{subsection}{Section}{Sections}

\definecolor{codebackground}{gray}{0.95}
\definecolor{codecomment}{rgb}{0.1,0.5,0.1}
\definecolor{codekeyword}{rgb}{0.2039,0.5804,0.7294}

\lstdefinelanguage{docker}{
  morekeywords={FROM, RUN, COPY, ADD, ENTRYPOINT, CMD,  ENV, ARG, WORKDIR, EXPOSE, LABEL, USER, VOLUME, STOPSIGNAL, ONBUILD, MAINTAINER},
  sensitive=true,
  morecomment=[l]{\#},
  morestring=[b]',
  morestring=[b]",
  showstringspaces=false
}

\lstdefinestyle{dfile}{
    backgroundcolor=\color{codebackground},
    breaklines=true,
    basicstyle=\ttfamily\footnotesize,
    commentstyle=\color{codecomment},
    keywordstyle=\color{codekeyword},
    frame=single,
    tabsize=2,
    language=docker
}

\lstdefinelanguage{yaml}{
  morekeywords = {schemaVersion, commandTests, name, command, args, expectedOutput, expectedError, fileExistenceTests, path, shouldExist, isExecutableBy, metadataTest, cmd, fileContentTests, expectedContents},
  sensitive=true,
  morecomment=[l]{\#},
  morestring=[b]',
  morestring=[b]",
  showstringspaces=false
}

\lstdefinestyle{cst}{
  backgroundcolor=\color{codebackground},
  breaklines=true,
  basicstyle=\ttfamily\footnotesize,
  commentstyle=\color{codecomment},
  keywordstyle=\color{codekeyword},
  frame=single,
  tabsize=2,
  language=yaml
}

\newcommand{\meta}{\textsf{metadataTest}}
\newcommand{\command}{\textsf{commandTests}}
\newcommand{\existence}{\textsf{fileExistenceTests}}
\newcommand{\content}{\textsf{fileContentTests}}

\setcopyright{acmlicensed}
\copyrightyear{2025}
\acmYear{2025}
\acmDOI{XXXXXXX.XXXXXXX}

\acmConference[EASE 2025]{The 29th International Conference 
on Evaluation and Assessment in Software Engineering}{17–20 June, 2025}{Istanbul, Turkey}
\acmISBN{978-X-XXXX-XXXX-X/YYYY/MM}

\begin{document}

\title{Toward Automated Test Generation for Dockerfiles\\
Based on Analysis of Docker Image Layers}

\author{Yuki Goto}
\affiliation{%
  \institution{Osaka University}
  \city{Suita}
  \state{Osaka}
  \country{Japan}
}
\email{yu-gotou@ist.osaka-u.ac.jp}

\author{Shinsuke Matsumoto}
\affiliation{%
  \institution{Osaka University}
  \city{Suita}
  \state{Osaka}
  \country{Japan}
}
\email{shinsuke@ist.osaka-u.ac.jp}

\author{Shinji Kusumoto}
\affiliation{%
  \institution{Osaka University}
  \city{Suita}
  \state{Osaka}
  \country{Japan}
}
\email{kusumoto.shinji.ist@osaka-u.ac.jp}

\begin{abstract}
  Docker has gained attention as a lightweight container-based virtualization platform.
  The process for building a Docker image is defined in a text file called a Dockerfile.
  A Dockerfile can be considered as a kind of source code that contains instructions on how to build a Docker image.
  Its behavior should be verified through testing, as is done for source code in a general programming language.
  For source code in languages such as Java, search-based test generation techniques have been proposed.
  However, existing automated test generation techniques cannot be applied to Dockerfiles.
  Since a Dockerfile does not contain branches, the coverage metric, typically used as an objective function in existing methods, becomes meaningless.
  In this study, we propose an automated test generation method for Dockerfiles based on processing results rather than processing steps.
  The proposed method determines which files should be tested and generates the corresponding tests based on an analysis of Dockerfile instructions and Docker image layers.
  The experimental results show that the proposed method can reproduce over 80\% of the tests created by developers.
\end{abstract}




\maketitle

\section{Introduction}
Docker has gained attention as a lightweight container-based virtualization platform.
Container-based virtualization enables the execution of applications without additional software installation in the local environment.
This approach not only enhances reproducibility and portability but also enables rapid deployment \cite{Sharma2016}.
Due to these advantages, Docker is widely used in software development \cite{Wu2023}.
Moreover, it has been leveraged as a way to ensure reproducibility in academic research \cite{Boettiger2015}.
The process for building a Docker image is defined in a text file called a Dockerfile.
A Docker image is built from this Dockerfile and a virtualized environment called a container is built from the Docker image.

A Dockerfile can be considered as a kind of source code that contains instructions on how to build a Docker image.
The behavior of Dockerfiles should be verified through testing, as is done for source code in a general programming language.
In Dockerfiles, instructions that require network communication are frequently executed.
These include package installation instructions via package managers and downloads using the \texttt{wget} or \texttt{curl} command.
Consequently, system degradation is more likely to occur due to external factors rather than the contents of a Dockerfile itself \cite{Henkel2021}.
Henkel et al. found that 26\% of the examined Dockerfiles on GitHub failed to build \cite{Henkel2021}.
Most of these failures were caused by changes in external environments, such as changes in dependency resolution.
Creating tests for a Dockerfile can also be useful for preserving behavior during Dockerfile refactoring.

Numerous automated test generation techniques have been proposed to support the testing of software written in a general programming language \cite{Anand2013, McMinn2004}.
Automatically generated tests are used for degradation prevention and refactoring.
Test generation techniques involve providing various input values for the arguments of methods and focusing on execution paths to maximize coverage.
For instance, EvoSuite \cite{Fraser2011}, a search-based test generation tool for Java, is commonly used.
EvoSuite can generate tests with high coverage.
Fraser and Arcuri reported that EvoSuite achieved 71\% branch coverage per class in large-scale empirical experiments \cite{Fraser2014}.
Test generation for Dockerfiles could also provide developer support.

However, existing search-based automated test generation techniques cannot be applied to Dockerfiles because a Dockerfile does not contain branches, only a single execution path.
As a result, the coverage metric, which is typically used as an objective function in existing search-based testing, becomes meaningless.
Therefore, the fundamental concept of exploratory testing cannot be applied.
A different approach is thus required.

In this study, we propose an automated test generation method for Dockerfiles.
The key concept of the proposed method is to generate tests based on processing results, rather than processing steps.
A Docker image built from a Dockerfile is considered as a set of effects produced by the Dockerfile.
Based on an analysis of Dockerfile instructions and Docker image layers, it is determined which files should be tested and the corresponding tests are generated.
Experimental evaluation confirms that the proposed method can reproduce over 80\% of the tests created by developers.

\section{Preliminaries}
\subsection{Dockerfile and Layers}
\label{sec:dockerfile_and_layers}
\begin{figure}[t]
  \centering
  \begin{subfigure}[b]{0.5\linewidth}
    \centering
\begin{lstlisting}[style=dfile]
# Set the base image
FROM debian:latest
# Install Python
RUN apt-get update && \
    apt-get install -y \
    python3
# Copy main.py
COPY main.py .
# Execute the script
# when starting a container
CMD ["python3", "main.py"]
\end{lstlisting}
    \subcaption{Example of Dockerfile}
    \label{fig:dfile-sample}
  \end{subfigure}
  \begin{subfigure}[b]{0.49\linewidth}
    \centering
    \raisebox{0.25\height}{
      \includegraphics[width=0.9\linewidth]{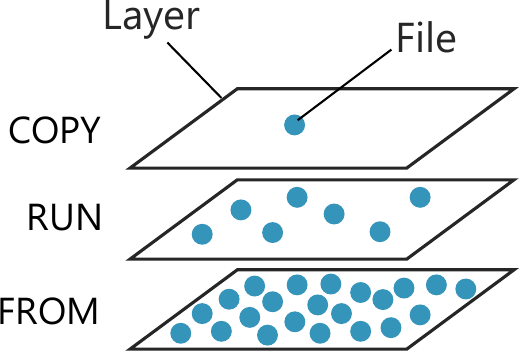}
    }
    \subcaption{Built Docker image}
    \label{fig:dimage-sample}
  \end{subfigure}
  \caption{Build execution example}
  \label{fig:build-sample}
\end{figure}

A Dockerfile is a text file that describes the steps required to build a Docker image.
\cref{fig:dfile-sample} shows an example of a Dockerfile that contains instructions to set up a Python runtime environment and execute \texttt{main.py}.
First, the FROM instruction specifies the base image, which serves as the foundation of the Docker image.
Next, the RUN instruction executes shell commands to install Python using a package manager.
Then, the COPY instruction copies the local \texttt{main.py} file into the image.
Finally, the CMD instruction sets the shell command to be executed when the container starts.
As shown in \cref{fig:dfile-sample}, multiple shell commands can be executed in the RUN section by using the shell operator \texttt{\&\&}.
\cref{fig:dimage-sample} shows the Docker image built from the Dockerfile in \cref{fig:dfile-sample}.
A Docker image can be considered as a snapshot of the configured environment.
A container, which is a virtual environment, is created from a Docker image.

A Docker image is composed of layers.
Each layer corresponds to an instruction in a Dockerfile.
During the build process, Dockerfile instructions are executed sequentially from top to bottom.
Each instruction generates a new layer that stores added, changed, and removed files.
For the Dockerfile shown in \cref{fig:dfile-sample}, the build process begins with the execution of the FROM instruction. 
As shown in \cref{fig:dimage-sample}, the layer created by the FROM instruction contains a large number of files required by the base image.
Next, when the RUN instruction is executed to install Python, a new layer is added.
This includes the \texttt{python3} binary along with related files and cached files generated by \texttt{apt-get}.
Following this, the COPY instruction creates an additional layer, where the \texttt{main.py} file is added.
Finally, the CMD instruction sets metadata for the image without changing the filesystem, so no additional layer is created.
These layers are transparently overlaid to form a unified file system.

\subsection{Container Structure Tests}
\label{sec:cst}
Container Structure Tests (CST) \footnote{\url{https://github.com/GoogleContainerTools/container-structure-test}} is a testing framework for Dockerfiles.
It can be used to verify the output of shell commands in a container and the existence of files.
The framework supports six types of test:
\begin{itemize}
  \item \textbf{\command{}: }Verify output of a shell command
  \item \textbf{\existence{}: }Confirm existence of a file
  \item \textbf{\content{}: }Inspect contents of a file
  \item \textbf{\meta{}: }Verify Docker image's metadata
  \item \textbf{\textsf{licenseTests}: }Check copyright files
  \item \textbf{\textsf{globalEnvVars}: }Verify environment variables
\end{itemize}
\cref{fig:cst-sample} shows an example of a test using CST for the Dockerfile shown in \cref{fig:dfile-sample}.
Here, we refer to a CST test unit as simply a \textit{test case}.
\cref{fig:cst-sample} shows three test cases.
The first test case, belonging to \command{}, verifies whether the Python runtime was correctly installed by running \texttt{python3 -{-}version} and checking that the version is 3.11.
The second test case, belonging to \existence{}, confirms the existence of \texttt{main.py} added by the COPY instruction in the Docker image's filesystem.
The third test case, belonging to \meta{}, ensures that the shell command specified by the CMD instruction is correctly configured.
In this study, CST is employed for testing Dockerfiles.

\begin{figure}[t]
  \centering
\begin{lstlisting}[style=cst]
commandTests:
  - name: 'check python3'
    command: 'python3'
    args: ['--version']
    expectedOutput: ['3\.11\..*']
fileExistenceTests:
  - name: 'check main.py'
    path: 'main.py'
    shouldExist: true
metadataTest:
  cmd: ['python3', 'main.py']
\end{lstlisting}
  \caption{Example of test using CST for Dockerfile shown in \cref{fig:dfile-sample}}
  \label{fig:cst-sample}
\end{figure}

\subsection{Automated Test Generation and Related Challenges}
Automated test generation is a technique that generates unit tests from source code in a bottom-up manner.
Various automated test generation tools have been proposed for different programming languages.
For example, EvoSuite \cite{Fraser2011} was proposed for Java and Pynguin \cite{Lukasczyk2022} was proposed for Python.
The goal of test generation is to cover as many execution paths as possible based on search.
For instance, EvoSuite adopts a genetic algorithm to search various input data to maximize coverage.
Search-based test generation assumes that the sufficiency of test cases can be measured by code coverage.

However, existing search-based test generation techniques are not applicable to Dockerfiles because a Dockerfile does not contain conditional branches, only a single execution path.
As a result, coverage as an objective function becomes meaningless.
Thus, the search process, which is fundamental to automated test generation, cannot be applied.
A different approach is thus required for automated test generation for Dockerfiles.

\section{Proposed Method}
\subsection{Overview}
\begin{figure}[t]
  \includegraphics[width=0.97\linewidth]{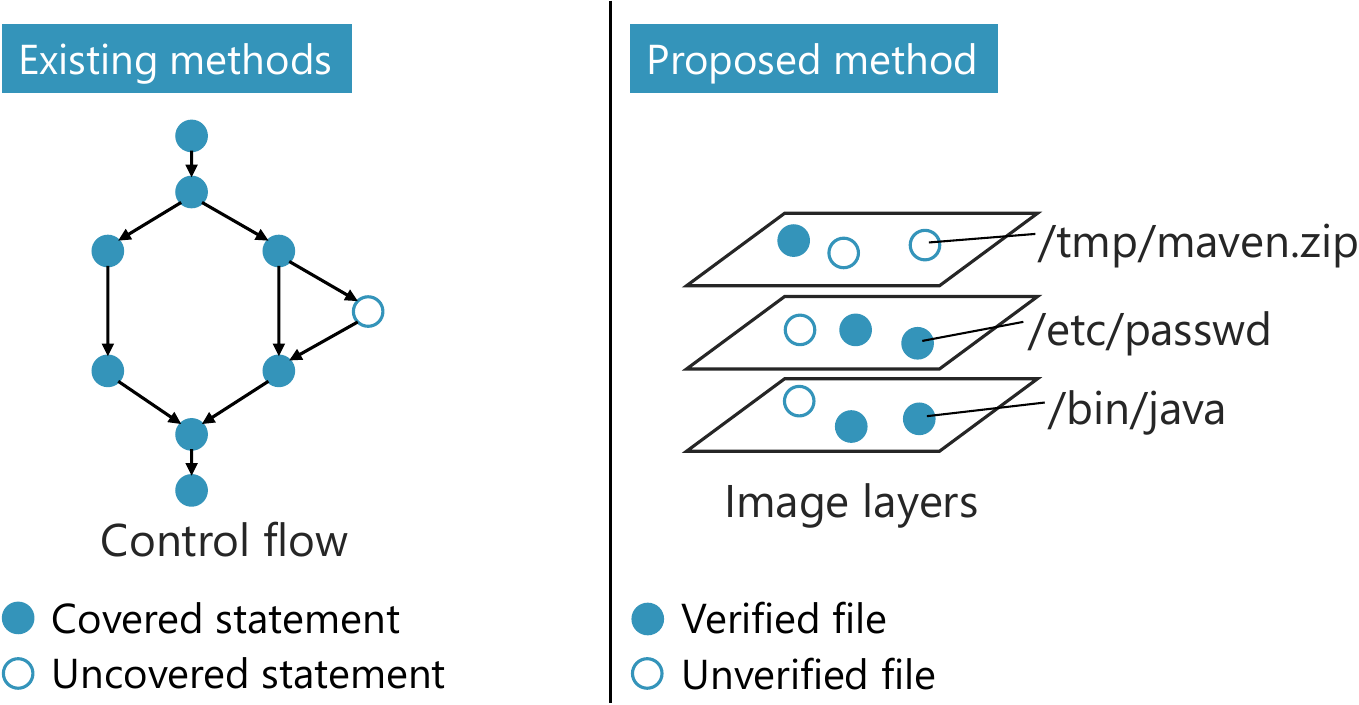}
  \centering
  \caption{Concept of proposed automated test generation}
  \label{fig:idea}
\end{figure}

In this study, we propose an automated test generation method for Dockerfiles to provide support for Docker developers.
The key concept of our method is to generate tests based on the execution results rather than execution procedures.
\cref{fig:idea} shows the difference between existing automated test generation and the proposed approach.
Existing approaches for general programming languages focus on coverage based on execution procedures.
In contrast, the proposed method attempts to generate tests based on execution results, specifically from a Docker image.
There is a similar concept in test quality evaluation, where a set of program effects is defined and the proportion of effects validated by assertions is regarded as a measure of test quality \cite{Koster2007, Schuler2013}.

As explained in \cref{sec:dockerfile_and_layers}, each layer of a Docker image corresponds to an instruction in a Dockerfile.
Thus, files in a layer can be considered as a collection of effects produced by Dockerfile instructions.
Verifying the match between these effects and the expected results allows sufficient confirmation of a Dockerfile's behavior.
However, tests that verify all effects not only have a high execution cost but also become sensitive to degradation.
Automatically generated tests tend to be more fragile than manually created tests \cite{Fewster1999}.
Therefore, the proposed method selects test targets based on importance, which is determined based on the Dockerfile instructions.
Finally, tests are generated in CST format (see \cref{sec:cst}).

\subsection{Procedure}
\begin{figure}[t]
  \includegraphics[width=0.8\linewidth]{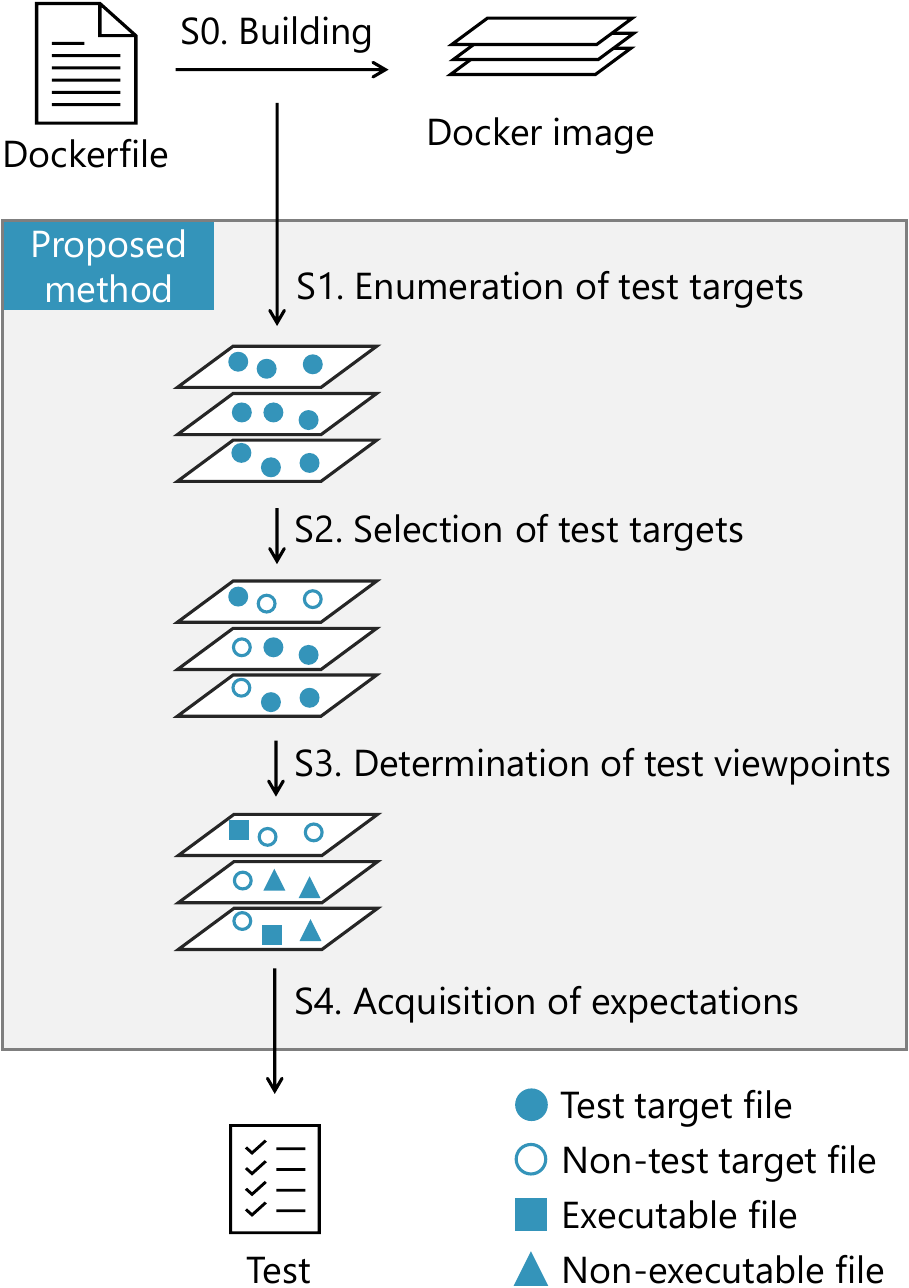}
  \centering
  \caption{Flow of test generation by proposed method}
  \label{fig:overview}
\end{figure}

\cref{fig:overview} shows the overall procedure of automated test generation by the proposed method.
The method consists of the next five steps.

\subsubsection*{S0. Building}
In this step, a Docker image is built from a Dockerfile.
In a Dockerfile, multiple shell commands can be written in the RUN section by using \texttt{\&\&}.
In the proposed method, the RUN section is split into multiple instructions before the Docker image is built.
This operation is used to break down the effects within layers and improve the accuracy of selecting test targets.

\subsubsection*{S1. Enumeration of Test Targets}
The effects of the Dockerfile are enumerated from the Docker image build in S0.
These effects are categorized into two types: metadata of the Docker image itself and files within the image layers.
Metadata are obtained using the \texttt{docker inspect} command, which provides detailed information about the Docker image.
Files are enumerated by extracting all files from each layer of the Docker image. 

Additionally, Dockerfile instructions are recorded.
The recording covers eight instructions that configure metadata, such as CMD, as well as four instructions that create layers: FROM, ADD, COPY, and RUN.
For layer-creating instructions, their arguments and other relevant information are linked to the corresponding layers.
The recording results are used in the next step to select test targets.

\subsubsection*{S2. Selection of Test Targets}
The test targets enumerated in S1 are selected based on the Dockerfile instructions.
This step aims to reduce the test execution cost and test fragility by limiting the Dockerfile effects.
The number of effects in a layer varies significantly depending on the instruction that created it.
For instance, a layer created by a COPY instruction that copies a single file contains only one file.
In contrast, a layer corresponding to a FROM instruction or a RUN instruction that installs packages may contain thousands of files.
Generating test cases for all such effects would result in a sufficiently comprehensive but excessive test set that checks unnecessary effects.
To address this issue, the effects that should be tested are determined based on their importance.

For this selection, we set heuristic-based scoring rules.
The scoring rules consist of 2 rules for metadata and 18 rules for files.
The metadata scoring rules are shown in \cref{tab:meta-scoring-rules} and the file scoring rules are shown in \cref{tab:file-scoring-rules}.
We created these rules based on investigations of developer-created tests using CST in 10 projects on GitHub.
We assumed that the tested targets reflected what developers considered important.
The rules were designed based on both tested and untested targets, so that frequently tested targets receive higher scores while untested ones receive lower scores.
After scores are assigned to all effects, only effects whose scores exceed a threshold are considered for test generation.
The number of generated tests can be adjusted by changing this threshold.

\begin{table}[b]
  \centering
  \caption{Scoring rules for metadata}
  \label{tab:meta-scoring-rules}
  {
  \tabcolsep = 4pt
  \begin{tabular}[b]{lr} \hline
  Condition & Score \\
  \hline
  Set by Dockerfile instruction & 10 \\
  Obtained from \texttt{docker inspect} command & 8 \\
  \hline
  \end{tabular}
  }
\end{table}
\begin{table}[b]
  \centering
  \caption{Scoring rules for files}
  \label{tab:file-scoring-rules}
  {
  \tabcolsep = 4pt
  \begin{tabular}[b]{lr} \hline
  Condition & Score \\
  \hline
  Path matches ADD or COPY destination & 9 \\
  Located under ADD or COPY destination directory & 3 \\
  Filename matches arguments of a command in RUN & 5 \\
  Path includes arguments of a command in RUN & 2 \\
  Path includes base image name or keyword& 3 \\
  Set as a working directory & 3 \\
  Set in environment variables & 2 \\
  Located under directories listed in \texttt{PATH} & 2\\
  Path includes \texttt{/bin/} & 3 \\
  Path includes \texttt{/etc/} & 3 \\
  Path includes \texttt{/conf/} & 3 \\
  Filename ends with \texttt{.sh} & 3 \\
  Generated by FROM & -5 \\
  Deleted & -10 \\
  Path starts with \texttt{/var/lib/apt/lists/} & -10 \\
  Path includes \texttt{/tmp/} & -10 \\
  Path includes \texttt{/cache/} & -10 \\
  Path includes \texttt{/log/} & -10 \\
  \hline
  \end{tabular}
  }
\end{table}

\subsubsection*{S3. Determination of Test Viewpoints}
In this step, the test viewpoints for the effects selected in S2 are determined.
The correctness of the metadata configuration is verified using the CST's \meta{}.
The approach for testing files differs between executable and non-executable files due to their distinct properties.
For executable files, \command{} are used to verify existence using a command such as \texttt{which} and check the version using options.
On the other hand, for non-executable files, \existence{} are used to check existence along with the absolute path.
These tests were determined by examining the tests created by developers, similar to the scoring rules of S2.
Note that executable files are defined as those with execution permissions within directories specified in the environment variable \texttt{PATH}.
Non-executable files are defined as all other files.

\subsubsection*{S4. Acquisition of Expectations}
To generate tests based on the test viewpoints defined in S3, expectations need to be obtained.
Similar to existing automated test generation, the expectations are determined in a bottom-up manner.
When generating \command{}, the execution of commands on a container is required to obtain the expectations.
To obtain the expectations for existence verification, the \texttt{which} command is run in the container with the executable's filename.
The output is then checked.
If the executable's path matches the output, an existence verification test case is generated.
The process then moves on to obtain the expectations for version verification.
If the executable's path differs from the output, a test case is generated as part of \existence{} rather than \command{}.
To obtain the expectations for version verification, a command with the option \texttt{-{-}version}, \texttt{-version}, or \texttt{-V} is executed in the container.
This checks the executable's version.
Once the output contains a string assumed to be the version, a test case is generated using the option and detected version.
In contrast, a test case using \meta{} or \existence{} can obtain expectations from the Dockerfile or Docker image.
Therefore, execution in the container is not needed.

\section{Evaluation}
\subsection{Overview}
To examine the performance of the proposed method, the following two experiments were conducted:

\noindent\textbf{Experiment 1:} Investigation of sufficiency of generated tests

\noindent\textbf{Experiment 2:} Investigation into whether developer-created tests can be reproduced by proposed method

The objective of Experiment 1 was to measure coverage, which indicates whether the proposed method can generate sufficient tests.
This was assessed by examining whether test cases were produced for all test targets.
In this experiment, we considered the generated tests to be sufficient if the test cases covered more than 95\% of files within image layers.
The objective of experiment 2 was to investigate how many test cases equivalent to those created by developers can be generated for each threshold.
This experiment can also be regarded as an evaluation of the effectiveness of the filtering mechanism in S2.

For the experiment, 10 open-source software projects that utilize CST for testing a Dockerfile were used.
An overview of these projects is shown in \cref{tab:target-project}.
To ensure diversity, each project had a distinct primary contributor.
If a project contained multiple Dockerfiles, only one was selected for the experiments.
Projects that used CST were used to allow comparison of the automatically generated tests with manually created tests in Experiment 2.
The developers created 25 \meta{}\textsf{s}, 34 \command{}, 47 \existence{}, and 14 \content{}, for a total of 120 test cases.
Typically, a \meta{} is counted as a single test case that verifies multiple items of metadata.
However, in the experiments, the number of \meta{}\textsf{s} was counted based on the number of metadata items verified.
The source code of the proposed method, details of the projects used in the experiments, and the experimental results are available at \url{https://zenodo.org/records/15023363}.

\begin{table}[b]
  \centering
  \caption{Open-source software projects used in experiments}
  \label{tab:target-project}
  {
  \tabcolsep = 4pt
  \begin{tabular}[b]{lrr} \hline
  Project name & Image size & \#Tests \\
  \hline
  zephinzer/cloudshell & 29.8 MB & 6 \\
  corretto/corretto-docker & 378 MB & 4 \\
  royge/deployer & 407 MB & 8 \\
  appwrite/docker-base & 1.11 GB & 37 \\
  jenkins-infra/docker-builder & 2.11 GB & 36 \\
  drecom/docker-rockylinux-ruby & 1.12 GB & 3 \\
  airdock-io/docker-sonarqube-scanner & 108 MB & 6 \\
  googleapis/testing-infra-docker & 3.87 GB & 9 \\
  liatrio/knowledge-share-app & 125 MB & 5 \\
  sassoftware/viya4-iac-aws & 2.32 GB & 6 \\
  \hline
  \end{tabular}
  }
\end{table}

\subsection{Experiment 1: Sufficiency}
\subsubsection*{Method.}
The purpose of this experiment was to confirm the sufficiency of the automatically generated tests.
To achieve this, we examined whether test cases were generated for all files within a Docker image's layers.
This experiment was conducted without selecting test targets in S2.
First, tests were generated using the proposed method without performing S2 for all 10 open-source software projects.
Next, the proportion of non-executable files targeted by \existence{} was determined.
For executable files, the percentage targeted by \command{} was determined.

\subsubsection*{Results.}
\begin{figure}[t]
  \includegraphics[width=0.97\linewidth]{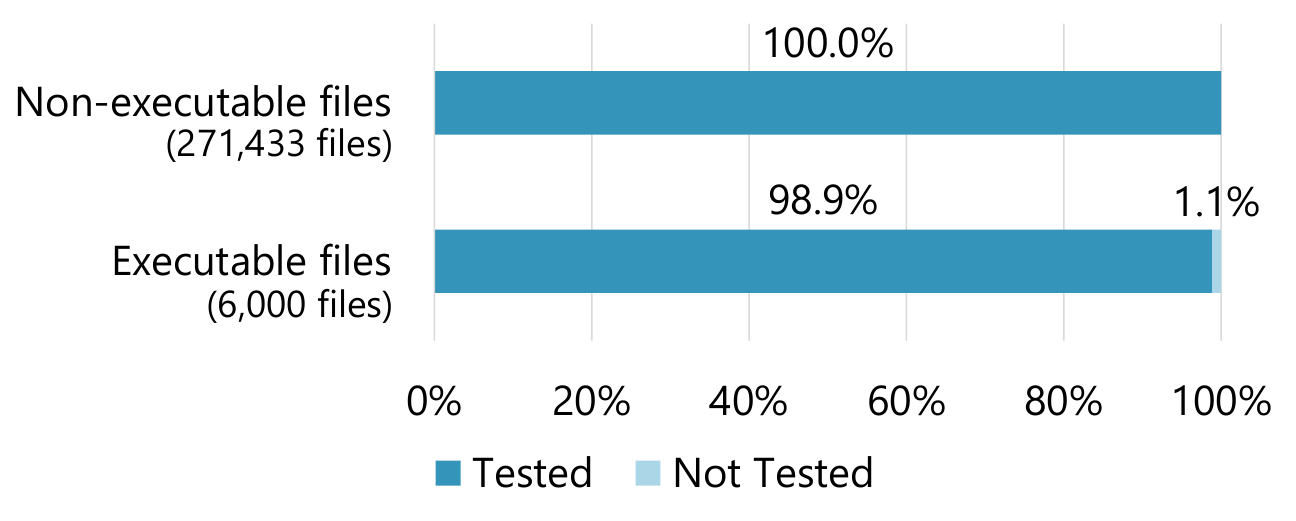}
  \centering
  \caption{Coverage of generated tests across all 10 projects}
  \label{fig:file-coverage}
\end{figure}

The results of Experiment 1 are shown in \cref{fig:file-coverage}.
As can be seen, \existence{} were generated for all non-executable files and \command{} were generated for most of the executable files.
As over 95\% of both file types were targeted, the generated tests were considered sufficient.
An examination of the generated test cases revealed that 49.7\% of the tested executable files had only their existence verified.
Among them, 95\% were executable files that did not provide any option to check their version.
The remaining executable files had version-check options, but their version verification test cases were not generated due to unsupported option formats or exit codes in the proposed method.
Since their behavior was not verified by executing them, these files should also be tested by running the commands.
On the other hand, 1.1\% of the executable files were not included in the test targets of \command{}.
These files were excluded because their location could not be confirmed using the \texttt{which} command in the container.
For such executable files, \existence{} were generated instead of \command{}.

\subsection{Experiment 2: Reproducing developer-created tests}
\subsubsection*{Method.}
Experiment 2 was used to investigate how many test cases equivalent to those created by developers can be generated for each threshold.
In other words, this experiment evaluated the selection of test targets in S2.
First, tests were generated using the proposed method for various thresholds and the number of generated test cases was counted.
Next, we manually verified how many of the generated test cases were equivalent to those created by developers and calculated precision and recall.
In this experiment, test cases that completely matched or had the same targets and testing intent, as shown in \cref{fig:same}, were considered equivalent.
We defined a complete match as a condition in which both the targets and assertions of the developer-created and the automatically generated test cases are identical.
Regarding the testing intent, test cases in \command{} that verify existence using a command such as \texttt{which}, as well as those in \existence{}, were classified as existence verification test cases.
Similarly, test cases in \command{} that check version using command line options were classified as version verification test cases.
When the classification of testing intent was same between developer-created and automatically generated test cases, they were considered to have the same intent.
Although some test cases in \existence{} might have included checks for execution permissions along with file existence, all such test cases were uniformly categorized as existence verification test cases.

Here, we discuss the results for the number of generated test cases using the project appwrite/docker-base.
This project was chosen as an example because its Docker image size is close to the average among those used in the experiments.
Note that the manually created tests were regarded as the ground truth in this experiment, even though reproducing them does not necessarily imply the ability to detect degradation.
\begin{figure}[t]
  \centering
  \begin{subfigure}[b]{0.97\linewidth}
    \centering
\begin{lstlisting}[style=cst]
fileExistenceTests:
  - name: "Make"
    path: "/usr/bin/make"
    shouldExist: true
    isExecutableBy: "any"
\end{lstlisting}
    \subcaption{Test case used to verify existence using \existence{}}
    \label{fig:same-dev}
    \vspace*{1.2ex}
  \end{subfigure}
  \begin{subfigure}[b]{0.97\linewidth}
    \centering
\begin{lstlisting}[style=cst]
commandTests:
  - name: 'check make'
    command: 'which'
    args: ['make']
    expectedOutput: ['/usr/bin/make']
\end{lstlisting}
    \subcaption{Test case used to verify existence using \command{}}
    \label{fig:same-pro}
  \end{subfigure}
  \caption{Example of test cases regarded as equivalent}
  \label{fig:same}
\end{figure}

\subsubsection*{Results.}
\begin{figure}[t]
  \includegraphics[width=0.97\linewidth]{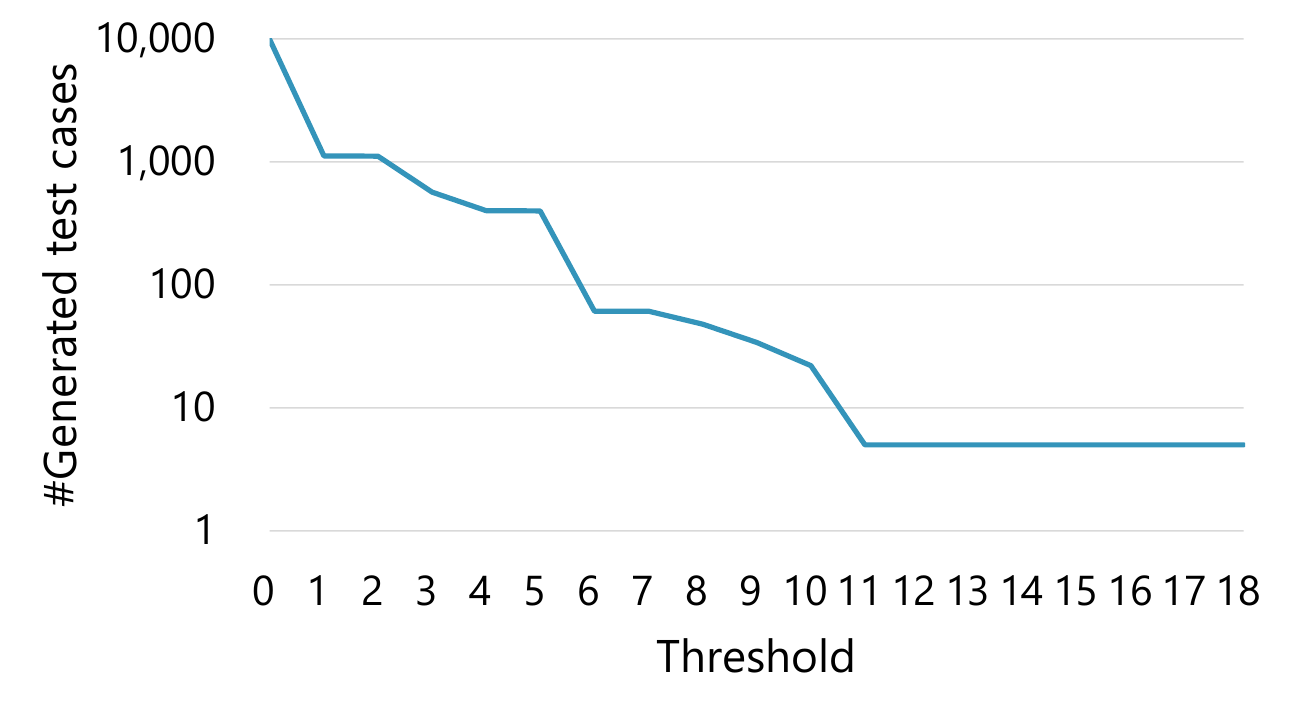}
  \centering
  \caption{Number of generated test cases for appwrite/docker-base}
  \label{fig:generation-graph}
\end{figure}

The results of applying the proposed method to the project appwrite/docker-base are shown in \cref{fig:generation-graph}.
The Dockerfile builds a Docker image with a size of 1.11 GB and 14,735 files.
We confirmed that adjusting the threshold allowed the number of generated test cases to be tuned from approximately 10,000 to just a few.
Similarly, the number of test cases could be reduced to a dozen or just a few for the other Dockerfiles.

\begin{figure}[t]
  \includegraphics[width=0.97\linewidth]{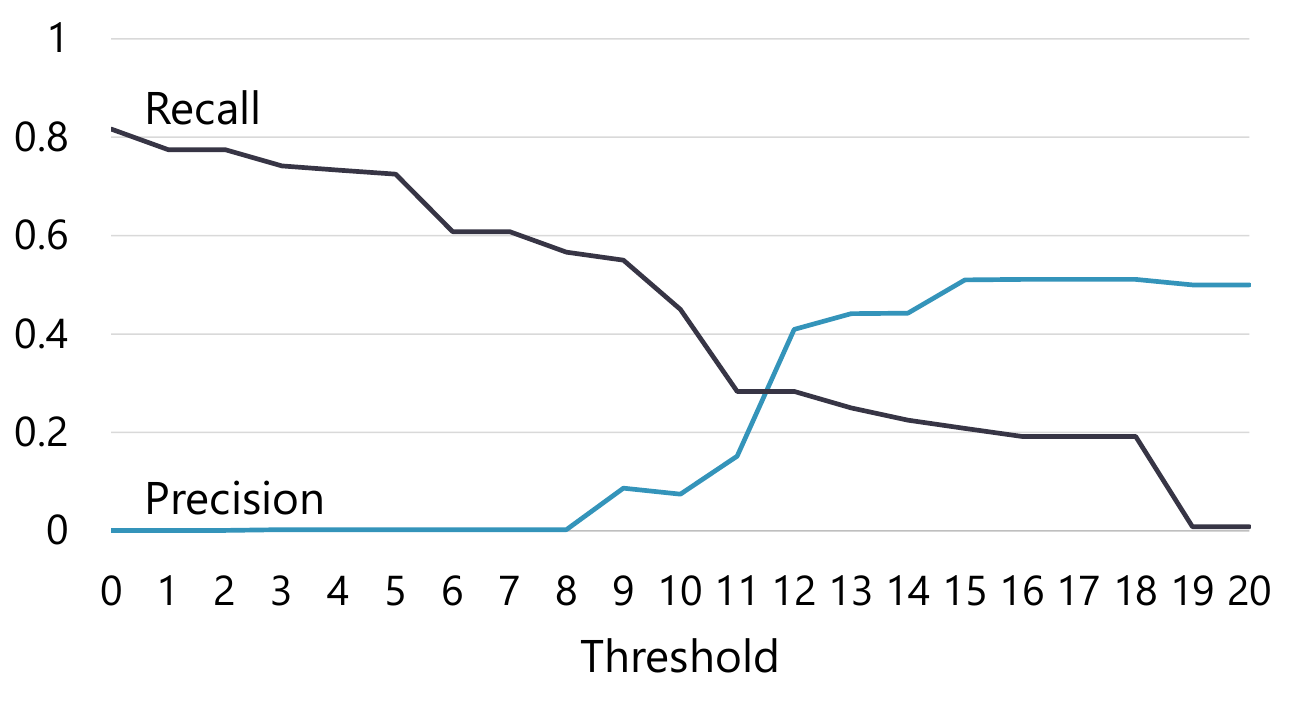}
  \centering
  \caption{Precision and recall of automatically generated tests compared to manually created tests for all 10 projects}
  \label{fig:pr-all}
\end{figure}

\cref{fig:pr-all} shows the results of the comparison of the tests generated by the proposed method with those created by developers.
The overall results for all 10 projects used in the experiment are shown.
The recall for a threshold of 0 shows that the proposed method reproduced over 80\% of the developer-created tests.
However, the precision at this threshold was nearly 0, indicating an excessive amount of generated test cases.
At the intersection of the precision and recall curves, the recall was approximately 0.3.
This issue is likely due to a problem with the scoring rules in S2.
To achieve higher recall, it is necessary to improve the rules.

\begin{figure}[t]
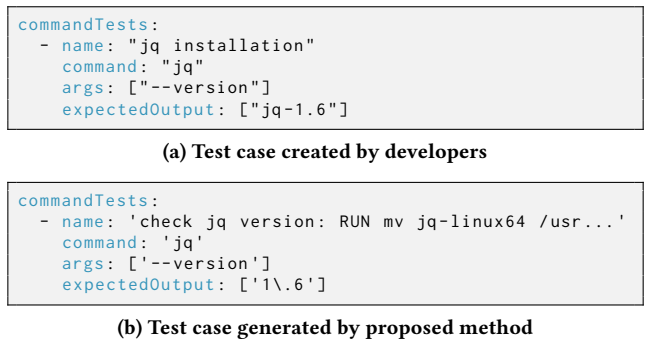

  \centering
  \begin{subfigure}[b]{0.97\linewidth}
    \centering
\begin{lstlisting}[style=cst]
commandTests:
  - name: "jq installation"
    command: "jq"
    args: ["--version"]
    expectedOutput: ["jq-1.6"]
\end{lstlisting}
    \subcaption{Test case created by developers}
    \label{fig:success-dev}
    \vspace*{1.2ex}
  \end{subfigure}
  \begin{subfigure}[b]{0.97\linewidth}
    \centering
\begin{lstlisting}[style=cst]
commandTests:
  - name: 'check jq version: RUN mv jq-linux64 /usr...'
    command: 'jq'
    args: ['--version']
    expectedOutput: ['1\.6']
\end{lstlisting}
    \subcaption{Test case generated by proposed method}
    \label{fig:success-pro}
  \end{subfigure}
  \caption{Example of equivalent test cases}
  \label{fig:success}
\end{figure}

An example where the proposed method generated test cases equivalent to those created by developers is shown in \cref{fig:success}.
\cref{fig:success-dev} was created by the developer and \cref{fig:success-pro} was generated by the proposed method.
Both of these test cases check the version of the \texttt{jq} command and thus have the same intention.

On the other hand, 19 test cases created by developers could not be generated by the proposed method.
An analysis of these test cases revealed that they could be classified into the following three categories:
\begin{enumerate}
  \item File contents verification: 14 test cases
  \item Detailed command information verification: 4 test cases
  \item Integration-test-like verification: 1 test case
\end{enumerate}
At present, the contents of files are not included in the scope of our method.
Thus, test cases that verify file contents were not generated.
Since such test cases account for about 10\% of the total, their generation will be a priority in future work.
An example of test cases that verify detailed information of a command is checking for extension modules.
This is a rare case and thus should not be prioritized in future work.
Integration tests that verify both an environment variable and a command version together were also not generated by the proposed method.
Since the proposed method generates unit tests, test cases that verify each component individually were generated.


\section{Conclusion and Future Work}
In this study, we proposed an automated Dockerfile test generation method.
This method is based on the processing results rather than the processing steps.
Through evaluation experiments, it was confirmed that the proposed method can generate test cases for most effects of Dockerfiles.
It was also found that the method can cover more than 80\% of the test cases created by developers.

In future work, two significant challenges need to be addressed.
The first challenge is to improve the proposed method.
A specific improvement is the expansion of the scoring rules to enhance the accuracy of test selection.
In addition, the generation of test cases that verify the contents of files should be added.
The second challenge is to investigate the effectiveness of the tests generated by the proposed method.
Although we conducted the experiment using developer-created tests in this paper, this is insufficient to evaluate the effectiveness of the generated tests.
Therefore, it is necessary to verify the tests' ability to detect degradation.
Furthermore, the excessiveness, or fragility, of the tests should be examined.

\begin{acks}
This research was partially supported by JSPS KAKENHI Japan 
(Grant Number: JP24H00692, JP21H04877, and JP21K18302) 
\end{acks}

\bibliographystyle{ACM-Reference-Format}
\bibliography{draft}


\begin{thebibliography}{12}


\ifx \showCODEN    \undefined \def \showCODEN     #1{\unskip}     \fi
\ifx \showISBNx    \undefined \def \showISBNx     #1{\unskip}     \fi
\ifx \showISBNxiii \undefined \def \showISBNxiii  #1{\unskip}     \fi
\ifx \showISSN     \undefined \def \showISSN      #1{\unskip}     \fi
\ifx \showLCCN     \undefined \def \showLCCN      #1{\unskip}     \fi
\ifx \shownote     \undefined \def \shownote      #1{#1}          \fi
\ifx \showarticletitle \undefined \def \showarticletitle #1{#1}   \fi
\ifx \showURL      \undefined \def \showURL       {\relax}        \fi
\providecommand\bibfield[2]{#2}
\providecommand\bibinfo[2]{#2}
\providecommand\natexlab[1]{#1}
\providecommand\showeprint[2][]{arXiv:#2}

\bibitem[Anand et~al\mbox{.}(2013)]%
        {Anand2013}
\bibfield{author}{\bibinfo{person}{Saswat Anand}, \bibinfo{person}{Edmund~K
  Burke}, \bibinfo{person}{Tsong~Yueh Chen}, \bibinfo{person}{John Clark},
  \bibinfo{person}{Myra~B Cohen}, \bibinfo{person}{Wolfgang Grieskamp},
  \bibinfo{person}{Mark Harman}, \bibinfo{person}{Mary~Jean Harrold},
  \bibinfo{person}{Phil McMinn}, \bibinfo{person}{Antonia Bertolino},
  {et~al\mbox{.}}} \bibinfo{year}{2013}\natexlab{}.
\newblock \showarticletitle{An orchestrated survey of methodologies for
  automated software test case generation}.
\newblock \bibinfo{journal}{\emph{Journal of systems and software}}
  \bibinfo{volume}{86}, \bibinfo{number}{8} (\bibinfo{year}{2013}),
  \bibinfo{pages}{1978--2001}.
\newblock


\bibitem[Boettiger(2015)]%
        {Boettiger2015}
\bibfield{author}{\bibinfo{person}{Carl Boettiger}.}
  \bibinfo{year}{2015}\natexlab{}.
\newblock \showarticletitle{An introduction to {Docker} for reproducible
  research}.
\newblock \bibinfo{journal}{\emph{Journal on Operating Systems Review}}
  \bibinfo{volume}{49}, \bibinfo{number}{1} (\bibinfo{year}{2015}),
  \bibinfo{pages}{71--79}.
\newblock


\bibitem[Fewster and Graham(1999)]%
        {Fewster1999}
\bibfield{author}{\bibinfo{person}{Mark Fewster} {and} \bibinfo{person}{Dorothy
  Graham}.} \bibinfo{year}{1999}\natexlab{}.
\newblock \bibinfo{booktitle}{\emph{Software Test automation: Effective use of
  test execution tools}}.
\newblock \bibinfo{publisher}{Addision-Wesley}.
\newblock


\bibitem[Fraser and Arcuri(2011)]%
        {Fraser2011}
\bibfield{author}{\bibinfo{person}{Gordon Fraser} {and} \bibinfo{person}{Andrea
  Arcuri}.} \bibinfo{year}{2011}\natexlab{}.
\newblock \showarticletitle{{EvoSuite}: automatic test suite generation for
  object-oriented software}. In \bibinfo{booktitle}{\emph{Proceedings of
  Symposium and European Conference on Foundations of Software Engineering}}.
  \bibinfo{pages}{416--419}.
\newblock


\bibitem[Fraser and Arcuri(2014)]%
        {Fraser2014}
\bibfield{author}{\bibinfo{person}{Gordon Fraser} {and} \bibinfo{person}{Andrea
  Arcuri}.} \bibinfo{year}{2014}\natexlab{}.
\newblock \showarticletitle{A large-scale evaluation of automated unit test
  generation using {EvoSuite}}.
\newblock \bibinfo{journal}{\emph{Transactions on Software Engineering and
  Methodology}} \bibinfo{volume}{24}, \bibinfo{number}{2}
  (\bibinfo{year}{2014}), \bibinfo{pages}{1--42}.
\newblock


\bibitem[Henkel et~al\mbox{.}(2021)]%
        {Henkel2021}
\bibfield{author}{\bibinfo{person}{Jordan Henkel}, \bibinfo{person}{Denini
  Silva}, \bibinfo{person}{Leopoldo Teixeira}, \bibinfo{person}{Marcelo
  d'Amorim}, {and} \bibinfo{person}{Thomas Reps}.}
  \bibinfo{year}{2021}\natexlab{}.
\newblock \showarticletitle{{Shipwright}: A human-in-the-loop system for
  {Dockerfile} repair}. In \bibinfo{booktitle}{\emph{Proceedings of
  International Conference on Software Engineering}}.
  \bibinfo{pages}{1148--1160}.
\newblock


\bibitem[Koster and Kao(2007)]%
        {Koster2007}
\bibfield{author}{\bibinfo{person}{Kenneth Koster} {and}
  \bibinfo{person}{David~C. Kao}.} \bibinfo{year}{2007}\natexlab{}.
\newblock \showarticletitle{State coverage: A structural test adequacy
  criterion for behavior checking}. In \bibinfo{booktitle}{\emph{Proceedings of
  Joint Meeting of the European Software Engineering Conference and Symposium
  on The Foundations of Software Engineering}}. \bibinfo{pages}{541--544}.
\newblock


\bibitem[Lukasczyk and Fraser(2022)]%
        {Lukasczyk2022}
\bibfield{author}{\bibinfo{person}{Stephan Lukasczyk} {and}
  \bibinfo{person}{Gordon Fraser}.} \bibinfo{year}{2022}\natexlab{}.
\newblock \showarticletitle{{Pynguin}: automated unit test generation for
  {Python}}. In \bibinfo{booktitle}{\emph{Proceedings of International
  Conference on Software Engineering}}. \bibinfo{pages}{168--172}.
\newblock


\bibitem[McMinn(2004)]%
        {McMinn2004}
\bibfield{author}{\bibinfo{person}{Phil McMinn}.}
  \bibinfo{year}{2004}\natexlab{}.
\newblock \showarticletitle{Search-based software test data generation: a
  survey}.
\newblock \bibinfo{journal}{\emph{Journal on Software testing, Verification and
  reliability}} \bibinfo{volume}{14}, \bibinfo{number}{2}
  (\bibinfo{year}{2004}), \bibinfo{pages}{105--156}.
\newblock


\bibitem[Schuler and Zeller(2013)]%
        {Schuler2013}
\bibfield{author}{\bibinfo{person}{David Schuler} {and}
  \bibinfo{person}{Andreas Zeller}.} \bibinfo{year}{2013}\natexlab{}.
\newblock \showarticletitle{Checked coverage: an indicator for oracle quality}.
\newblock \bibinfo{journal}{\emph{Journal on Software Testing, Verification and
  Reliability}} \bibinfo{volume}{23}, \bibinfo{number}{7}
  (\bibinfo{year}{2013}), \bibinfo{pages}{531--551}.
\newblock


\bibitem[Sharma et~al\mbox{.}(2016)]%
        {Sharma2016}
\bibfield{author}{\bibinfo{person}{Prateek Sharma}, \bibinfo{person}{Lucas
  Chaufournier}, \bibinfo{person}{Prashant Shenoy}, {and}
  \bibinfo{person}{Y.~C. Tay}.} \bibinfo{year}{2016}\natexlab{}.
\newblock \showarticletitle{Containers and virtual machines at scale: A
  comparative study}. In \bibinfo{booktitle}{\emph{Proceedings of International
  Middleware Conference}}. \bibinfo{pages}{1--13}.
\newblock


\bibitem[Wu et~al\mbox{.}(2023)]%
        {Wu2023}
\bibfield{author}{\bibinfo{person}{Yiwen Wu}, \bibinfo{person}{Yang Zhang},
  \bibinfo{person}{Kele Xu}, \bibinfo{person}{Tao Wang}, {and}
  \bibinfo{person}{Huaimin Wang}.} \bibinfo{year}{2023}\natexlab{}.
\newblock \showarticletitle{Understanding and predicting {Docker} build
  duration: An empirical study of containerized workflow of {OSS} projects}. In
  \bibinfo{booktitle}{\emph{Proceedings of International Conference on
  Automated Software Engineering}}. \bibinfo{pages}{1--13}.
\newblock


\end{thebibliography}

\end{document}